\documentclass{aa} 
\usepackage{graphicx}
 
\begin{document}

\thesaurus{03        
           (19.92.1  
            19.94.1  
            07.09.1  
            07.22.1) 
                   }         
\title{A new determination of supernova rates and a comparison with 
indicators for galactic star formation.}
\titlerunning{New supernova rates and galactic star formation}
\author{E.\ Cappellaro\inst{1} \and R. Evans\inst{2} \and M.\ Turatto\inst{1}}
\institute{Osservatorio Astronomico di Padova, vicolo dell'Osservatorio 5,
I-35122 Padova, Italy
\and
P.O. Box 131, Hazelbrook NSW 2779, Australia
}

\offprints{E. Cappellaro}

\date{Received ................; accepted ................}

\maketitle

\begin{abstract}

We have computed new estimates of the local rates of supernovae (SNe)
adding the updated log of Evans' visual search to our SN search
database. In this way, we have accumulated the largest SN statistics ever
assembled for this purpose.

The new SN rates are corrected on an empirical basis for the bias in
the inner regions of galaxies and that in inclined spirals. We also
tested an alternative approach based on the simple model proposed by
Hatano et al. (1998) for the SN and dust distribution in spirals. It
turns out that, although the two approaches give similar average
rates, the Hatano et al. model appears to overcorrect the SN rate of
distant galaxies.

We used these updated statistics to probe the SN rates with different
tracers of the star formation activity in galaxies, namely integrated
colors, infrared luminosities and nuclear activities.  We found a
clear relation between the core-collapse SN rate and the integrated
galaxy color, which appears consistent with the prediction of galaxy
evolutionary models.  We also compared SN rates in galaxies with
different $L_{FIR}$ with unfavorable outcome, and we argue that
$L_{FIR}$ is not a universal measurement of SFR.

Finally, we confirm that the SN rate is not enhanced in AGN host
galaxies which indicates that the nuclear engine does not
significantly stimulate the extranuclear SF.

\end{abstract}


\section{Introduction}

One of the frontiers of current astronomical research is the
observation of supernovae at high-redshift. In fact it is expected
that by using SNe~Ia as distance indicators it will be possible to
constrain the geometry of the Universe within a few years.  Equally
important and difficult, is to determine the rate of the different
types of SNe as a function of redshift. The rationale is that the
various types of SNe have progenitors of different ages; in particular
core-collapse SN~II+Ib/c result from young, massive stars and SN~Ia
originate from intermediate to old population stars (eg. Branch et
al. \cite{bnf}). Therefore, the evolution of the relative SN rates
with redshift can be used to probe the average SFR rate history in
galaxies and, in turn, constrain scenarios for galaxy formation and
evolution.

So far, there have been only exploratory attempts in this direction
(J{\o}rgensen et al. \cite{jorg}; Sadat et al. \cite{sadat}; Madau et
al. \cite{madau}), which however have demonstrated the potential of this
approach and motivated new observational efforts.

The accurate determination of the present time SN rates is the
benchmark, crucial to exploiting these efforts to the full. Another
important factor is to compare the rates of various SN types with
different indicators of the stellar population content of galaxies
in the local Universe.
 
One problem with these rates is that local SNe are rare and therefore it
requires several years or decades to collect sufficient statistics.
In addition, in order to obtain accurate estimates of the SN
rate, it is necessary to know: $i)$ the sample of galaxies which have been
searched for SNe, $ii)$ the frequency and limiting magnitude of
observations and $iii)$ the instruments/techniques which are used for
detection in order to assess search biases. 

Very few groups of professional astronomers have had the perseverance
and force to carry out a SN search program long enough to be really
useful for this purpose ({\em cf} Cappellaro et al. \cite{stat95}
hereafter C97).  Among the amateurs in this field, an outstanding case
is the visual SN search which has been conducted by Evans since 1980
(Evans \cite{ev:97}).  Indeed, estimates of the SN rate based on the
first 10 years of Evans' SN search have already been published (Evans et
al. \cite{ev:89}; van den Bergh
\& Mc~Clure \cite{vdbmc}). In this paper we will analyze the updated log of this survey which doubles the statistics with respect to previously
published estimates (Sec.~\ref{evans}).

Following the protocol described in a previous paper (C97), we pooled
together Evans' log and those of  photographic searches and used
the improved statistical basis to test a different approach for the
correction of selection effects (Sec.~\ref{sec_hatano}).

We used this combined database, which is the largest ever built for
such a purpose, to obtain updated estimates of the SN rates
(Sec.~\ref{update}).  Finally, we compared the SN rates with other SFR
indicators such as the far infrared luminosity, the integrated
color and the activity of the galaxy.

\section{Evans' visual SN search}\label{evans}

Evans began his search of SNe in 1980 using a 25~cm telescope.  The
observations were conducted visually, which has the advantage of being
very fast and inexpensive, but the limiting magnitude for SN
discovery is not very deep ($m_{\rm lim}=14.5$ mag).  At the end of
1985 the telescope was replaced with a 41~cm telescope ($m_{\rm
lim}= 15.0$ mag), and most recently complemented by a 100~cm
telescope at Siding Spring Observatory ($m_{\rm lim}=16.0$).

\begin{table*}
\caption{SN rates in spirals of different inclination (not corrected 
for the inclination bias.)}\label{inc}
\begin{tabular}{cccccccc}
\hline
       & \multicolumn{3}{c}{Evans search} &&
       \multicolumn{3}{c}{photographic searches (from C97)} \\
\cline{2-4}\cline{6-8}
inc           &   N.    &   N. &  rate     && N.    & N. &  rate  \\
 ${\rm [deg]}$& galaxies  & SNe & [SNu]$^*$ &&galaxies &SNe&[SNu]$^*$\\
\hline     
0-45     &   616  & 18 &  $0.88\pm0.21$ && 1581 & 36 & $0.94\pm0.16$ \\   
45-65    &   666  & 13 &  $0.49\pm0.14$ && 1702 & 25 & $0.59\pm0.12$ \\    
65-90    &   592  & 11 &  $0.34\pm0.10$ && 1818 & 15 & $0.32\pm0.08$\\ 
\hline
\end{tabular}

$*$ $1 \,{\rm SNu} = 1\, {\rm SN}\, (100 {\rm yr})^{-1}\, (10^{10}
L_\odot^{\rm B})^{-1}$. \\
 
\end{table*}

During the almost two decades of the search, Evans collected over
200,000 individual observations, surveying a sample of more than 3000
nearby galaxies. The search resulted in the discovery of 32 SNe, and
another 22 SNe, which first had been discovered by others were also
detected, for a total sample of 54 SNe.  These numbers qualify Evans'
search as the most successful amateur SN search ever and make it very
competitive even to professional searches.

For the calculation of SN rates based on Evans' search log, we
used the control time method and the protocol described in C97.  The
essential galaxy data, that is distances, morphological types,
luminosities and axial ratios, have been retrieved from the updated
version of the RC3 catalog (de Vaucouleurs et al. \cite{rc3})
distributed by the ``Centre de Donn{\'e}es Astronomiques de
Strasbourg''. With the improved SN statistics, the Poissonian
errors on SN events were reduced and other sources of errors
became dominant, in particular in the correction
of search biases.

All SN searches suffer from specific biases and therefore it is
important to compare the results of different kinds of searches.
Because it is unique, Evans' visual search is especially useful for
comparisons with traditional photographic surveys (C97).  Specific
biases in SN searches occur because SNe appear embedded in their
parent galaxies.  Since the detection efficiency of new objects
depends on the contrast against the background, obviously it is more
difficult to discover a SN in the (luminous) inner regions of a galaxy
than in its outskirts.  This bias is at its most pronounced if one
uses a wide field/small scale telescope and a detector with a small
dynamic range. Also the fraction of SNe lost is greater in more
distant galaxies because of the small angular size of the galaxy
image.  By comparing the radial distributions of SNe in galaxies at
different distances, we estimated that up to 50\% of the SNe exploding
in the more distant galaxies are lost in photographic searches using
Schmidt telescopes. The nuclear bias seems negligible only in CCD and
visual searches of nearby galaxies (C97).

A precursory examination of the general list of SNe reveals another
severe bias against SN detection: that in inclined spirals. In the
past, it has been claimed that this bias does not affect visual and
CCD searches (Evans et al. \cite{ev:89}; Muller et al. \cite{muller})
but so far the evidence has been inconclusive (C97).  It is thus of
interest to exploit the improved statistics of Evans' visual
search to address this question.

We began by calculating the overall SN rate in spiral galaxies of
different inclination both for Evans' visual search and for the
combined photographic search sample constructed by C97, including the
SN searches of Asiago (Cappellaro et al. \cite{PI}), Crimea (Tsvetkov
\cite{tsv:83}), OCA (Pollas
\cite{pol}) and Cal\'an/Tololo (Hamuy et al. \cite{mario:93}).
For this particular calculation we turned off the correction for
parent galaxy inclination which was included in the recipe of C97.
The results are reported in Table~\ref{inc}, where the bins were
chosen to give roughly the same number of galaxies in each inclination
bin.  It would appear that based only on Evans' log, the SN rate
in edge-on spirals is 2.6 fold less than in face-on ones. This
clearly demonstrates that the bias in inclined spirals also affects 
visual searches and is almost as severe as in photographic searches
(2.9).

The natural interpretation of this bias is that SNe occurring in the
disk of inclined spirals appear on the average dimmer than those in
face-on spirals because of the increased optical depth through the
dust layer. As a consequence, the probability of SN discovery in
inclined spirals is reduced. In line with this interpretation, it is
expected that the bias is more severe in searches carried out in the
blue band, (e.g. photographic searches), than in visual ones (CCD
searches in the red would be even less affected).

As a first order approach to correct for the inclination bias, we can
assume a plane parallel geometry for the distribution of dust in the
disk of spirals. In this case the average extinction of the SN
population scales with $\sec i$, where $i$ is the inclination of the
galaxy disk with respect to the line of sight. We showed in C97
that this assumption results in an over-correction of the SN rate in
edge-on galaxies.  It was argued that this is evidence that dust is
not uniformly distributed in the disk of spirals but is instead in
discrete clouds.  Until more evidence is available we have adopted, in
analogy to C97, a conservative extinction law that is intermediate
between the $\sec i$ and an empirical relation, derived from the
assumption that the SN rate is the same in face-on and edge-on
galaxies (see Section ~\ref{sec_hatano} for an alternative model).

When such a correction is included, we can compute the SN rates for
the complete Evans search and compare it with those obtained from the
first ten years of the search and with those derived from the combined
photographic search sample.  The results are reported in
Table~\ref{evares} where the number of galaxies and SNe for each
sample is indicated in the first two rows.  We must stress that, since
we adopt the same galaxy catalog, input parameters, bias corrections
and numerical recipe, the differences between the columns in
Table~\ref{evares} are only due to the different logs of observations.

The new Evans rates are consistent with the earlier results but, at
least with respect to the average value, in much better agreement with
the rate from photographic searches. Looking more in detail, it
appears that for early type galaxies (E-S0 and S0a-Sb) the updated
Evans value is in better agreement with the photographic search rate,
whereas for late spirals (Sbc-Sd) the old value was closer.  We attribute
these fluctuations to the small statistics of individual SN searches,
which become wider when the sample is divided into bins.

\begin{table}
\caption{Comparison of the SN rate [SNu] obtained from the Evans'
updated statistics (1980-1998), the first 10 years of the search
(1980-1989) and the combined photographic search sample}
\label{evares}
\begin{tabular}{lccc}
\hline
           & Evans 80-98 & Evans 80-89 & ph. search \\
N. galaxies & 3068   & 1377  &  7319 \\
N. SNe       & 54  & 24  & 94  \\
\hline
\\
E-S0   & $0.17\pm0.06$ & $0.13\pm0.08$ & $0.18\pm0.05$ \\
S0a-Sb & $0.83\pm0.20$ & $1.22\pm0.41$ & $0.66\pm0.13$  \\
Sbc-Sd & $0.96\pm0.13$ & $1.20\pm0.38$ & $1.34\pm0.20$ \\
\\
All$^*$& $0.66\pm0.09$ & $0.81\pm0.17$ & $0.67\pm0.07$  \\
\hline
\end{tabular}

$*$ Including Sm, irregulars and peculiars.
\end{table}

\section{An alternative model for bias corrections}\label{sec_hatano}

As mentioned before, corrections for search biases are the most
controversial step in the calculation of SN rates. In our approach
(cf. C97), the correction factors are tuned to cancel the sign of the
biases from the calculated SN rates regardless of their physical
causes. The ideal would be to build a model for the dust and SN
distribution which is consistent with the present data on galaxies and
SN progenitor populations and derive from it estimates of the
biases.

In a recent paper, Hatano et al. (\cite{hatano}) described a simple
model based on an assumed distribution of the dust and SN populations
which predicts that, because of extinction, SNe in inclined spirals
appear on average dimmer and shoe a much wider magnitude scatter than
those in face-on spirals. Moreover, because the dust distribution
peaks in the central regions of galaxies, this effect is more
pronounced for SNe occurring in those regions.

This provides an alternative to the classical explanation of the
selection effect in the central region of galaxies which would thus
derive from the enhanced extinction of SNe instead of the reduced
luminosity contrast. According to this scenario, the bias would be
most severe in dust, inclined spirals and, because of the small scale
height, for core collapse SNe.  Indeed, all these features were
found in the observed SN sample (Table~1 of Cappellaro \& Turatto
\cite{ct:97}).

A special characteristic of the Hatano et al. model is that core
collapse SNe do not occur within 3 Kpc of the center of the galaxy. 
Therefore SN~II or Ib/c do not appear in the central regions of
face-on galaxies although an increasing number of core collapse SNe
appears to be projected on the centers of the more inclined spirals
due to projection effects. In any case, type Ia SNe are more highly
concentrated than type II or Ib/c.  Though Hatano et al. claim that
the observations confirm their model, we should mention that van den
Bergh (\cite{vdb97}) and Wang et al. (\cite{wang}) reached the
opposite conclusion on the basis of similar data.

Even if the Hatano et al. model should be considered as exploratory
given the above controversy, it is of interest to test how adopting it
can change the SN rate estimates.  We thus have replaced the empirical
bias corrections mentioned in Section~\ref{evans} with the observed SN
luminosity distribution for each SN type in spirals of different
inclination derived from the Hatano et al. (1998) model.  Then we
computed the control times for each bin of the luminosity function for
each galaxy and SN type. The total control time was obtained as the
weighted average according to the observed luminosity distribution .

The results of this calculation are shown in Table~\ref{hattab}.  Taken
at face value and compared with the empirical bias corrections
(Tab.~\ref{final}), we found that by using the Hatano et al. model,
the SN~Ia rate in the entire sample of spirals results 10-20\% higher
and the SN~II+Ib/c rate 15\% smaller. These differences all fall
within the errors and should not be considered significant.

\begin{table}
\caption{The SN rate corrected using the Hatano et al. (1998) model}\label{hattab}
\begin{tabular}{lccc}
\hline  
galaxy   &\multicolumn{3}{c}{rate [SNu]} \\
type    &  Ia   &   II+Ib/c & All\\
\hline
S0a-Sb  &  $0.27\pm0.08$ & $0.63\pm0.24$      & $0.91\pm0.26$\\
Sbc-Sd  &  $0.24\pm0.10$ & $0.86\pm0.31$      & $1.10\pm0.32$\\
\\	  				 
Spirals$^*$ &  $0.25\pm0.09$ & $0.76\pm0.27$  & $1.01\pm0.29$\\
\hline
\end{tabular}

$*$ Includes types from Sm, irregulars and peculiars.
\end{table}

We notice, however, that by adopting the Hatano et al. model the SN rate
in edge-on spirals remains 1.5 times smaller than in face-on spirals, and
that the rate in intermediate inclination spirals is even
smaller. Slightly increasing the optical depth of the dust layer 
helps but does not solve the problem.

A more perplexing feature of the bias corrections based on the Hatano
et al. model is that it produces SN rates which increase with galaxy
distances: the rate in galaxies with $v>3000$ km s$^{-1}$ results
almost twice that in galaxies with $v<3000$ km s$^{-1}$. This could be
resolved by reducing the average reddening but which is the opposite
of the previous recommendation.  This apparent contradiction may
simply indicate that, as already suggested, the SN progenitors and
dust distributions in real galaxies are more complex than in this
simple exploratory model.  In conclusion, though the approach seems
promising, the Hatano model needs further refinement and in the
meanwhile we decided to maintain the {\em empirical} bias corrections of C97.

\section{SN rates and indicators of the galactic SFR }\label{update}

Once we had verified that the SN rates derived from Evans' visual
search are similar to those obtained from photographic searches and
decided the correction for search biases, we merged all search logs in
a single database.  In this way we obtained an improvement in the
statistics compared to C97 (from 110 to 137 SNe) and equally important
we balanced the weights of different types of searches (over one third
of the SNe in the new sample were discovered in  Evans' visual
search).

The SN rates computed using these updated statistics are reported in
Table~\ref{final} \footnote{Through this paper we assumed H$_0$=75 km
s$^{-1}$ Mpc$^{-1}$. SN rates reported in this paper can be
transformed to other values of the Hubble constant multiplying by
(H$_0/75)^2$} where errors include not only event statistics but also
uncertainties in the input parameters and in the bias corrections. The
differences with C97 are small ($<15$\%) and well within the errors.
We notice that in C97 the SN~Ia rate appeared to increase when
progressing from early to late type galaxies whereas this effect had
now nearly vanished. The relatively low rates of SN~Ib/c compared with
SNII were, instead, confirmed.

\begin{table*}
\caption{SN rate( in SNu) from the combined search sample.}\label{final}
\begin{tabular}{lrrrccccc}
\hline  
galaxy  & \multicolumn{3}{c}{N. SNe$^*$} &&\multicolumn{3}{c}{rate [SNu]} \\
\cline{2-4}\cline{6-9}
type    & Ia   &   Ib/c &  II  &&  Ia   &   Ib/c &  II    & All\\
\hline
E-S0    & 22.0 &      &      && $0.18\pm0.06$ & $<0.01$       & $<0.02$       & $0.18\pm0.06$\\
S0a-Sb  & 18.5 &  5.5 & 16.0 && $0.18\pm0.07$ & $0.11\pm0.06$ & $0.42\pm0.19$ & $0.72\pm0.21$\\
Sbc-Sd  & 22.4 &  7.1 & 31.5 && $0.21\pm0.08$ & $0.14\pm0.07$ & $0.86\pm0.35$ & $1.21\pm0.37$\\
Others$^\#$  &  6.8 &  2.2 &  5.0 && $0.40\pm0.16$ & $0.22\pm0.16$ & $0.65\pm0.39$ & $1.26\pm0.45$\\
\\
All & 69.6 &  14.9& 52.5&& $0.20\pm0.06$ & $0.08\pm0.04$ & $0.40\pm0.19$ & $0.68\pm0.20$\\
\hline
\end{tabular}

$*$ Similar to C97, 10 unclassified SNe have been redistributed among
the three basic SN types according to the observed distribution that is
100\% Ia in E-S0, in spirals: type Ia 35\%, type Ib 15\%, type II 50\%.

$\#$ Others includes types Sm, Irregulars and Peculiars
\end{table*}

The normalization of the SN rate to the galaxy blue luminosity has
been introduced after the demonstration that the former scales with the
latter.  This is convenient because $i)$ integrated $B$ magnitudes are
available for a large number of galaxies and $ii)$ the B luminosity
for a given galaxy type scales with the total mass at a first
approximation.  Physically, the blue luminosity is a good tracer of
the young stellar population in starburst galaxies, but not in normal
galaxies where a considerable fraction of the continuum luminosities
is produced by old stars also in the blue (Sage \& Solomon
\cite{sage}; Kennicutt \cite{kenni}).

In principle, by using different photometric bands should be possible
to sample selected stellar populations and hence to obtain useful
information for progenitor scenarios.  For instance, van den Bergh
(\cite{vdb:90}) and Della Valle \& Livio (\cite{mdv:94}) normalized
the rate of SN~Ia to H and K luminosity: in these bands the role of
old stars in all galaxy types is dominant. If all SN~Ia result
from low mass stars we would expect the SN~Ia rate per unit of H and K
luminosities not to be correlated to galaxy type.  The fact that the
rate in these units increases considerably when moving from
ellipticals to late spirals was taken to indicate that a significant
fraction of SN~Ia result from intermediate age stars.  Even if their
conclusion is probably correct, it must be stressed that these
estimates were not direct measurements but a simple scaling of the
SN rates in unit blue luminosity based on the assumption of an average
B-H and B-K color per galaxy type.  Because H and K photometry is
available only for a small fraction of the galaxies in our sample,
unfortunately, the SN rate in these units cannot be directly measured.

It would also be of interest to estimate the
rates of SNe, in particular of core collapse SNe, in galaxies with
different star formation rates (SFR).

The different diagnostic methods which are used to probe SFR in
galaxies have been reviewed in a recent paper by Kennicutt
(\cite{kenni}). Because we are limited by the SN statistics we need
tracers that are available for large samples of galaxies. In this
respect, integrated colors and far infrared (FIR) luminosities are
particularly appealing.

\subsection{SN rates and galaxy integrated colors}\label{sec_color}

Integrated broad band colors are very useful for statistical purposes,
as they are reliable indicators of the galaxy stellar population with
bluer galaxies expected to host stars that are younger and more
massive than redder ones.  Colors are most interesting 
because, by using evolutionary synthesis models, it is possible to
estimate the SFR per unit mass or luminosity required to produce a
given integrated color for a given stellar population. It is well
known that along the Hubble sequence the galaxy color becomes bluer
moving from early to late types and that this corresponds to a
sequence in SFR which is virtually zero in ellipticals and maximum in
late spirals.  However, especially in spirals, there is a significant
dispersion in the average color from galaxy to galaxy, indicating
that SFR can vary significantly even for a given Hubble type.

Conveniently, $(B-V)_T^0$ and $(U-B)_T^0$ colors, corrected for
galactic and internal extinction are listed in the RC3 catalog for a
fair percentage of the galaxies of our sample (24\% and 19\%
respectively). From these we derived also $(U-V)_T^0$ colors which,
allowing for the extended wavelength baseline, are more sensitive SFR
indicators.  For each bin of galaxy morphological types we divided the
galaxies into subsamples, containing galaxies bluer and redder than
the global average. We then computed separately the SN rates for each
of these subsamples.  The results are reported in Table~\ref{color},
where the galaxy types are in col~1, the average colors for the
galaxies of the specific subset are in cols~2 and 5, the SN rates in
SNu for SN~Ia and for core-collapse SNII+Ib/c in cols~3-4 and 6-7.

\begin{table*}
\caption{SN rates in SNu for galaxies with integrated colors
bluer and redder than the average.
}\label{color}
\begin{tabular}{lccc|ccc}
\hline  
galaxy  & \multicolumn{3}{c|}{blue galaxies} &\multicolumn{3}{c}{red galaxies} \\
type    &  $<(B-V)_T^0>$ & Ia   &  II+Ib  & $<(B-V)_T^0>$ & Ia   &  II+Ib  \\
\hline
E-S0    & 0.86 & $0.3\pm0.1$ &             & 0.95 & $0.2\pm0.1$ \\
S0a-Sb  & 0.60 & $0.2\pm0.1$ & $0.6\pm0.2$ & 0.80 & $0.2\pm0.1$ & $0.5\pm0.2$\\
Sbc-Sd  & 0.45 & $0.1\pm0.1$ & $1.5\pm0.3$ & 0.62 & $0.3\pm0.1$ & $0.9\pm0.2$\\
        &      &             &             &      &             &\\
All$^*$ & 0.56 & $0.2\pm0.1$ & $1.0\pm0.2$ & 0.88 & $0.2\pm0.1$ & $0.1\pm0.1$\\
\hline
\end{tabular}

\begin{tabular}{lccc|ccc}
\hline  
galaxy  & \multicolumn{3}{c|}{blue galaxies} &\multicolumn{3}{c}{red galaxies} \\
type    &  $<(U-V)_T^0>$ & Ia   &  II+Ib  & $<(U-V)_T^0>$ & Ia   &  II+Ib  \\
\hline
E-S0    & 1.26 & $0.3\pm0.1$ &             & 1.48 & $0.2\pm0.1$ \\
S0a-Sb  & 0.67 & $0.2\pm0.1$ & $0.9\pm0.3$ & 1.12 & $0.2\pm0.1$ & $0.4\pm0.2$\\
Sbc-Sd  & 0.28 & $0.2\pm0.1$ & $1.7\pm0.5$ & 0.62 & $0.3\pm0.1$ & $0.8\pm0.3$\\
        &      &             &             &      &             &\\
All$^*$ & 0.54 & $0.2\pm0.1$ & $1.1\pm0.2$ & 1.32 & $0.2\pm0.1$ & $0.1\pm0.1$\\
\hline
\end{tabular}

$*$ Including Sm, irregulars and peculiars.
\end{table*}

As expected, the rate of core collapse SNe (II+Ib/c)
is higher in the bluer spirals. By using B-V color this effect is seen
only for late spirals (the rate is higher by a factor of 1.7 for Sbc-Sd),
but becomes clear for all spirals when using U-V color (over a factor
of 2). Instead the rate of SN~Ia is, within the uncertainties,
independent on galaxy colors.

With regards to the rows labeled ``All'' (which includes galaxies of
all types) we should note that dividing galaxies into bluer and redder
colors to a large extent corresponds to separating them into early
and late type galaxies.  Therefore the great difference in the core
collapse SN rates in bluer and redder galaxies simply reflects the
fact that core collapse SNe are not found in early type galaxies.

\begin{figure}
\resizebox{\hsize}{!}{\includegraphics*{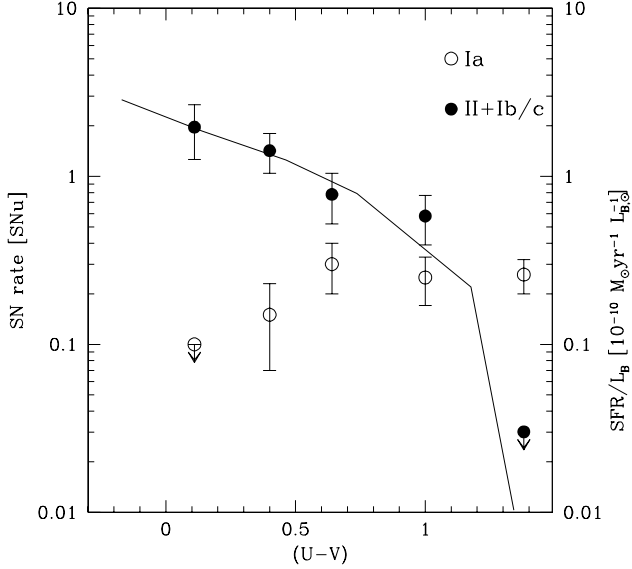}}
\caption{SN rates in SNu (left-hand scale) in spirals with different
U-V color (for this plot early and late spirals have been considered
together).  Filled symbols are core-collapse SN~II+Ib/c and open
circle are SN~Ia. Error-bars only accounts for the SN statistics.  The
line gives the SFR rate per unit B luminosity (right-hand scale) for
galaxies with different colors as predicted by the evolutionary
synthesis models of Kennicutt (1998). The surprising coincidence of
the two scales can be partially understood by simple units conversion
(see text).}\label{sfr}
\end{figure}

We can compare the observed SN rates with the predicted SFR in
galaxies of different colors. This is done in Fig.\ref{sfr} where the
dots represent the SN rates in SNu (left-hand scale) in galaxies of
different $U-V$ integrated colors and the line is the SFR per unit 
of blue luminosity (right-hand scale) taken  
from the evolutionary synthesis models of Kennicutt (\cite{kenni}).

In general, for a galaxy of luminosity L$_B$, because of the short
life of progenitor evolution, the number of core collapse SNe per
century corresponds to the number of new born stars within the
appropriate mass range, namely:

$$ SN\,rate [SNu] \times L_B \simeq {SFR \times f_{M_L}^{M_U} \over <M_{SN}>} \times 100 $$

where $f_{M_L}^{M_U}$ is the mass fraction of stars which are born
with mass in the range $M_L$ to $M_U$, the lower and upper limit for
core-collapse SN progenitors, and $<M_{SN}>$ is the
average mass of SN progenitors. According to the standard scenarios,
$M_L \simeq 8 M_\odot$ and $M_U \simeq 40 M_\odot$. 
Adopting a Salpeter mass function, $f_{M_L}^{M_U} \simeq 10^{-1}$ and
$<M_{SN}> \simeq 10 M_\odot$ which compensate the factor 100 which accounts 
for the difference in the time scale.

In conclusion, even if the exact coincidence of the two scales in our
figure is to some degree fortuitous, the nice agreement of the SFR
measured through core collapse SN rates and that deduced by synthesis
modeling for ``average'' spiral galaxies, lends support to the general
scenario for stellar population evolution.

Conversely, the fact that the rate of SN~Ia shows no dependence on the
galaxy U-V color requires a significant delay between the SFR episodes
and the onset of SN~Ia events.

The relation between core collapse SN rates and colors provides a
useful tool for the comparison of local and high-$z$ SN rates.  Indeed,
for galaxies at high-$z$ integrated colors can be measured relatively
easily, whereas morphological types, requiring superb imaging, are not
generally available.  Conversely, it is clear that reporting the
average SN rates for uncharacterized galaxy samples may
turn out to be pointless for constraining galaxy evolution models.

\subsection{SN rates and galaxy FIR luminosities}\label{sec_fir}

The interest in deriving the SN rate in units of the FIR
luminosities was stressed by J{\o}rgensen (\cite{jorg90}) who made a
first attempt based on the general SN catalog. Here we report our
calculations based on the control time technique.

The near infrared emission of spiral galaxies shows at least two
components: a warm component associated with dust around young stars
and a cool component associated with more extended dusty heated by the
general stellar radiation field, including radiation from old stars.
The warm emission gives a direct measurement of the SFR, but in normal
galaxies it is heavily contaminated by the cool component.

In addition to the extended star formation in the disk, many spiral
galaxies show an enhanced SFR in the nuclear region. The observations
show that the nuclear and extended components are mostly decoupled. In
these ``starburst'' galaxies the nuclear SFR could reach 1-1000
M$_\odot yr^{-1}$ and the integrated infrared emission is largely
dominated by the nuclear component (Kennicutt \cite{kenni}). At the
same time we expect a very high rate of core collapse SNe, which are
however difficult to detect in the optical. This is because in the
nuclear starburst regions one expects several magnitudes of extinction
and a severe bias for optical SN searches. These considerations must
be kept in mind when interpreting the results.

FIR fluxes have been measured by the IRAS survey for over 30000
galaxies in the range $10-100 \mu{\rm m}$ and FIR magnitudes are
reported in the RC3 catalog for $\sim 30$\% of the galaxies of our
sample. They have been converted to units of solar FIR luminosites
using the relation:

$$ {L_{\rm FIR} \over L_{\rm FIR,\odot}} = 10^{-0.4 m_{\rm FIR}}\, 3.1
\times 10^{11}\, d^2 $$

where $d$ is the galaxy distance in Mpc.

First we computed the SN rates for unit infrared luminosity
$L_{\odot,{\rm FIR}}$.

\begin{table}
\caption{SN rates per unit FIR luminosity. $ 1\,{\rm SNuIR} = 1\, {\rm SN}
(100 yr)^{-1} (10^{10} L_{{\rm FIR},\odot})^{-1}. $ }
\label{lfir}
\begin{tabular}{lcccc}
\hline  
galaxy  &\multicolumn{3}{c}{SN rate [SNuIR]} \\
type    &  Ia   &   II+Ib/c    & All\\
\hline 
E-S0    & $1.8\pm0.8$ &             & $1.8\pm0.8$\\
S0a-Sb  & $0.6\pm0.2$ & $2.0\pm0.5$ & $2.7\pm0.5$\\
Sbc-Sd  & $0.6\pm0.1$ & $3.5\pm0.6$ & $4.1\pm0.6$\\
\\     
All$^*$ & $0.7\pm0.1$ & $2.5\pm0.3$ & $3.2\pm0.3$\\
\hline   
\end{tabular}

$*$ Includes types Sm, Irregulars and Peculiars.
\end{table}

If the FIR luminosity is a direct measure of the SFR in spirals, as 
is often assumed, we would expect the rate of core collapse SNe
per unit FIR luminosity to be constant through all galaxy types.
Instead, the results reported in Table~\ref{lfir} show that the rate of
core collapse SNe in FIR units increases almost  2 fold moving
from early to late spirals, whereas the rate of SN~Ia remains constant
(in Table~\ref{lfir} we report the SN rate in SNuIR
also for E-S0 galaxies, though we do not expect there to be any relation
between L$_{\rm FIR}$ and SFR) . We have already stressed that there are
different contributing factors to FIR luminosities, and in particular
early spiral galaxies often exhibit low temperature, relatively high
FIR luminosities attributable to dust heating from the general
stellar radiation field, and not directly related to SFR (Kennicutt
\cite{kenni}).

It has been claimed that a more reliable discriminant of the SFR is
the infrared excess L$_{\rm FIR}$/L$_{\rm B}$ (eg. Tomita et
al. \cite{tomita}). This is because by normalizing to the blue
luminosity we partially remove the effect of the general radiation
field.  In Table~\ref{lfirlb} we report the SN rates in SNu for
galaxies with different infrared excess, along with that of galaxies
not detected by IRAS. Though in general we cannot translate ``not
detected'' into a precise upper limit, it is reasonable to assume
that, for our RC3 galaxy sample, the average FIR luminosity of the
undetected sample is smaller than that of the detected sample. Support
for this belief comes from the fact that the distance distributions of
the detected and not detected RC3 galaxy samples are similar.

\begin{table*}
\caption{SN rates in SNu for galaxies with different infrared excess
}\label{lfirlb}
\begin{tabular}{lcccccc}
\hline  
galaxy & \multicolumn{2}{c}{not detected by IRAS} &
\multicolumn{2}{c}{$L_{FIR}/L_{B}\le 0.35$} &
\multicolumn{2}{c}{$L_{FIR}/L_{B}>0.35$} \\ type & Ia & II+Ib/c & Ia &
II+Ib/c & Ia & II+Ib/c \\
\hline 
E-S0    &$0.2\pm0.1$&            & $0.4\pm0.2$ &             &$<0.5$      &\\
S0a-Sb  &$0.2\pm0.1$& $0.3\pm0.2$& $0.2\pm0.1$ & $0.5\pm0.2$ &$0.3\pm0.1$ &$1.1\pm0.4$\\
Scd-Sd  &$0.3\pm0.1$& $0.7\pm0.3$& $0.2\pm0.1$ & $1.0\pm0.2$ &$0.2\pm0.1$ &$1.2\pm0.3$\\
\\
All$^*$ &$0.2\pm0.1$& $0.2\pm0.1$& $0.2\pm0.1$ & $0.9\pm0.1$ &$0.3\pm0.1$ &$1.1\pm0.2$\\     
\hline   
\end{tabular}

$*$ Including Sm, irregulars and peculiars.
\end{table*}	

The rate of core collapse SNe is higher in the IR detected galaxies
compared with the not detected sample, whereas this is not the case
for SN~Ia (Table~\ref{lfirlb}) whilst there are no significant
differences between galaxies with small and large infrared excess.
This again supports the idea that, whereas a fraction of the FIR
luminosity originates from SF regions, the other contributing
factors to the IR emission of galaxies eliminate, at least in normal
galaxies, the relation between L$_{\rm FIR}$ and SFR.

\subsection{SN rates in active galaxies}\label{AGN}

It is generally believed that nuclear activity stimulates the SF
(Rodriguez-Espinoza et al. \cite{rodri}) and therefore that the rate
of core-collapse SNe in AGN must be higher than in normal galaxies.
An open issue is whether the SF is stimulated throughout the whole AGN host
galaxy or only in the circumnuclear region. From the observational
point of view, in the first case we would expect an enhanced detection rate,
whereas due to the high extinction in the nuclear starburst regions
this may not occur in the latter case.

To address this question we crossed our RC3 galaxy list with the
Catalog of Quasars and Active Galactic Nuclei of V\'eron-Cetty \&
V\'eron (\cite{veron}) (distributed by the CDS).
This catalog contains a list of almost 15000 quasars and AGN most of
which are too distant for normal SN searches (only $\sim
1100$ have recession velocities smaller than 15000 km s$^{-1}$).

We found that 283 galaxies out of our combined RC3 sample,  ($\sim$
3\%) are also listed in the V\`eron-Cetty \& V\`eron catalog (this simply
reflects the relative occurrence of AGN in the local Universe).  Most
of them ($\sim$ 88\%) are Seyfert and the rest HII galaxies.  In these
galaxies our searches have discovered 17 SNe, that is 12\% of the total
SN sample.  This could be taken as evidence that the SN rate in AGN
is enhanced compared with the general sample.

However, when the control time method is applied, the average SN rate in the
AGN sample is $0.6\pm0.1$ SNu ($0.4\pm0.1$ SNu for core-collapse SNe),
identical to that of the general sample ($0.7\pm0.1$
SNu for all SNe and $0.5\pm0.1$ SNu for core-collapses).  We note that
the AGN sample shows roughly the same distribution of morphological
types as the general sample.

There are two reasons why the high detection rate in our AGN galaxy
sample does not reflect in higher SN rates in SNu.  First of all, the
average control time for a galaxy of the AGN galaxy sample (5.08 yr)
is almost twice that of the general galaxy sample (2.67 yr). 
Secondly, the galaxies of the AGN sample are over 2 times more
luminous ($<L_B> = 3.1^{10} L_\odot$) than the average ``normal''
galaxies ($<L_B> = 1.4^{10} L_\odot$).  This stresses the risks of
interpreting  statistics derived from general SN samples and not
from actual search logs.

A similar conclusion was reached by Richmond et al. (\cite{richm}) as
the result of a dedicated SN search in 142 nearby starburst
galaxies. They obtained 1.1 SNu (scaled to $H_0=75$) for the
total rate and 0.7 for core-collapse only, somewhat larger than our
corresponding estimates.  However, their statistical error
is also quite large (they had a sample of only 5 SNe) and allowing
also for the different computational protocols, the difference should not
be regarded as significant.

The conclusion is that the SN rate in active galaxies is the same as
in normal ones (cf Petrosian \& Turatto \cite{petro}).  More
precisely, this finding only applies to the AGN host galaxies and not
to the AGNs themselves, which because of the high extinction rate
cannot be probed by optical SN searches. Therefore our claim is that
the nuclear engine does not significantly stimulate the SFR outside
the nuclear region of the host galaxy.

\section{Conclusions}

We have presented new estimates of the SN rates in galaxies, obtained
by including the updated log of the Evans' visual SN search in our
database. In this way we have obtained a sample of 137 SNe in a
reference sample of about $10^4$ galaxies. Based on the comparison
between visual and photographic surveys we tested the effectiveness of
the bias corrections and verified that they are consistent with our
understanding of galaxies and SN progenitors. In particular, we show
that the Hatano et al. (\cite{hatano}) simple model for the SN and
dust distributions in galaxies explains, at least to the first order, both
the bias in the nuclear region and in inclined spirals, though
actually some refinement is needed before it can be used to correct SN
rates.

The new rates have been compared with other tracers of the average SFR
in galaxies.  We found that the rates of core-collapse SNe are higher
in bluer spirals, while the same is not true for SNIa. This was
expected, since bluer galaxies host stars that are younger and more
massive than redder ones. In particular, we find that the correlation
between galaxy colors and core-collapse SN rates is similar to that
predicted by the evolutionary models of Kennicutt (\cite{kenni}).

We have found that there is not a direct relation between core-collapse SN
rates and FIR luminosities confirming that FIR luminosity is not a
universal measurement of SFR. This can be explained by considering
that FIR emission in galaxies at least in the normal ones, is made up of
different components and not exclusively related to young stars.

Finally, our data confirms previous findings that the SN rates in AGN
host galaxies are not enhanced. This conclusion does not apply to the
nuclear starburst regions which cannot be probed by current SN
searches.

\begin{acknowledgements}
R.E. wishes to thank Mr. R. McDonell of Ark Angles Business and
Personal Computer Systems, Hazelbrook (Australia) for designing the
computer program used to record his observations. We wish also to
thank Janet Clench for revising the manuscript.
\end{acknowledgements}

\bibliographystyle{astron}
\bibliography{/home/enrico/testi/bibliografia}

\end{document}